\documentclass[pre,twocolumn,aps,superscriptaddress,showpacs]{revtex4}
\usepackage{amsmath}
\usepackage{graphicx}

\begin{document}

\title{Testing simplified protein models of the hPin1 WW domain}

\author{Fabio Cecconi}         
\address{INFM-CNR, Istituto dei Sistemi Complessi (ISC) 
Via dei Taurini 19, I-00185, Rome, Italy}
\author{Carlo Guardiani}
\address{ 
Centro Interdipartimentale per lo Studio delle Dinamiche Complesse (CSDC) 
Sezione INFN di Firenze, (Italy)}
\author{Roberto Livi}
\address{Dipartimento di Fisica Universit\`a di Firenze \\
Centro Interdipartimentale per lo Studio delle Dinamiche Complesse (CSDC),\\
Sezione INFN di Firenze e INFM UdR Firenze, Italy.}



\begin{abstract}
The WW domain of the human Pin1 protein for its simple topology and
the large amount of experimental data is 
an ideal candidate to assess theoretical approaches to protein folding. 
The purpose of the present 
work is to compare the reliability of the chemically-based Sorenson/Head-Gordon 
(SHG) model and a standard native centric model in reproducing 
through  molecular dynamics simulations some of the 
well known features of the folding transition of this small domain. 
Our results show that the G\={o} model correctly reproduces the 
cooperative, two-state, folding mechanism of the WW-domain, 
while the SHG model predicts 
a transition occurring in two stages: a collapse followed by a structural 
rearrangement. The lack of a cooperative folding   
in the SHG simulations appears to be related to the non-funnel shape of the energy 
landscape featuring a partitioning of the native valley in sub-basins 
corresponding to different chain chiralities. 
However the SHG approach remains more reliable in estimating the   
$\Phi$-values with respect to G\={o}-like description. This may suggest 
that the WW-domain folding process is stirred by energetic
and topological factors as well, and it highlights the better suitability of 
chemically-based models in simulating mutations.
\end{abstract}


\maketitle

\section{Introduction}

The WW domains are a family of fast-folding, compact, modular domains 
featuring a triple-stranded, anti-parallel beta-sheet owing their name 
to the presence of two highly conserved Triptophanes (W).
Recent studies \citep{Ferguson} suggested that 
these domains may fold at a rate 
close to the speed limit for $\beta$-sheet formation offering the 
opportunity to investigate the pathways of $\beta$-sheet kinetics.
\citep{BrooksIII}. 
In particular, the 
human Pin1 protein WW domain, due to the availability of several 
structural~\citep{Verdecia, Bayer}, 
thermodynamical and kinetic~\citep{Gruebele} experimental data, represents an 
excellent target to test computational techniques and theoretical approaches.

\begin{figure}[!h]
   \begin{center}
      \includegraphics*[width=\columnwidth]{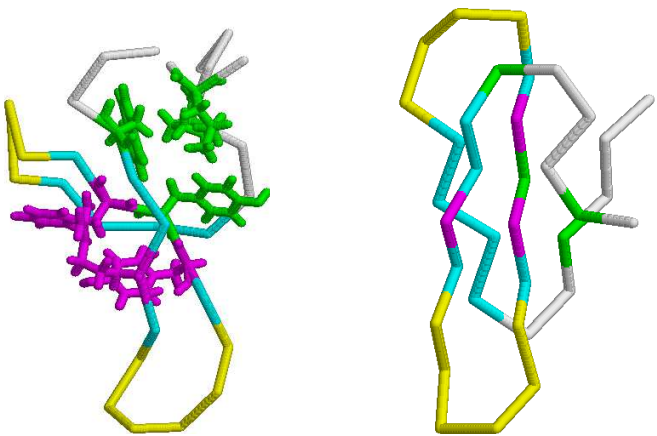}
   \end{center}
\caption{Backbone representation of the NMR structure of Pin1 WW 
domain  (pdb-id = 1NMV, Left) and simulated structure $\Gamma_0$ (Right). 
Residues in the three $\beta$-strands are 
coloured in blue, those belonging to loops L1 and L2 in yellow. 
Side-chains of residues participating in CL1 (Leu7, Trp11, Tyr24, Pro37) are 
shown in green stick representation, those involved in CL2 
(Arg14, Tyr23, Phe25) in magenta. 
Figures were drawn with RASMOL.}
\label{fig:struct}
\end{figure}

The structure of this domain, resolved through NMR~\citep{Bayer} 
and X-ray diffraction~\citep{Verdecia} (Fig. \ref{fig:struct}),  
is characterized by hydrophobic clusters providing the largest contribution 
to the thermodynamic stability~\citep{Gruebele}. 

Cluster 1 (CL1) involves residues 
Leu7, Trp11, Tyr24 and Pro37, the second cluster (CL2) comprises 
Tyr23, Phe25 and Arg14. 
The stability of the molecule also derives from a network of 
hydrogen-bonds whose central element is the highly conserved Asn26 
located on strand $\beta_2$ and 
acting both as donor and acceptor in bonds with Pro9, Trp11, 
Ile28 and Thr29, thus linking strands $\beta_1$ and $\beta_3$. 
The presence of two loops. 
Loop I (L1) plays a key role in substrate 
recognition~\citep{Verdecia} as it binds to the phosphate of the 
\mbox{pS-P} motif of the 
Proline-rich ligands, Loop II (L2), instead, gives an important contribution 
to thermal stability~\citep{Gruebele}.
Thermal denaturation experiments \citep{Gruebele} and simplified statistical 
physics approaches 
\cite{Bruscolini} have shown that the Pin1 WW domain folds following 
cooperative two-state mechanism at the temperature $T_M=332$ K.
The mutagenesis analysis performed by the same authors 
\citep{Gruebele} identified the mutations on Ser16, Ser18 and Ser19 in 
Loop I as maximally destabilizing for the transition state, so that the 
formation of L1 appears to be the 
rate-limiting step in the folding/unfolding process. Loop II (L2)
is involved in the formation of the transition state only at high 
temperatures~\citep{Gruebele}. 
Due to the ability of inducing 
conformational changes in Proline-rich, phosphorilated substrates, Pin1 is a potential regulator 
of the cell-cycle, and maybe involved in pathologies like Liddle's syndrome, muscular dystrophy 
and Alzheimer's disease~\citep{Pin1-pathol1, Pin1-pathol2}.

The aim of this work is the comparison of two off-lattice protein descriptions: 
the G\={o}-model \citep{Gomodel} which customary allows studying the influence of the 
native state structure on the folding process, and a model
proposed by Sorenson and Head-Gordon \citep{SHG1,SHG2} (hereafter referred to as SHG), 
mainly based on the primary and secondary structural information. 
The conceptual justification of topology-based or native centric models relies on the 
observation that the
topology of native states can play a crucial role in selecting some features of the 
folding mechanism \citep{Mich,Eaton,Baker,Fersht}. 
The main experimental finding supporting the above statement 
can be summarized as follows (Baker \citep{Baker2}): (a) 
the similarity shared by transition-state 
conformations and folding mechanisms of proteins having structurally related native 
states despite their low sequence homology \citep{Serrano,Serrano2,Riddle} and 
(b) the correlation that certain simple topological properties, such as contact order, 
may have 
with protein folding rates \citep{Chiti,Plaxco}. 
The G\={o}-force field is independent of the
amino acid sequence, and it requires the knowledge of the tertiary structure of 
native states to identify native interactions. Accordingly, 
the native centric approach cannot be used for ab initio predictions of 
native folds, even if recent works \citep{Clementi,Finkel,Zhou,Shakhnovich,Bahar} 
provide growing evidence that it can be confidently used for the 
characterization of transition states of real small, fast-folding 
(sub-millisecond) 
\citep{Fast,Fast1} proteins that are characterized by a low level of energetic
frustration. However, topological models might not correctly 
reproduce the folding process when chemical interactions play a relevant
role.  
The SHG model, which is instead based on the chemical and physical 
properties of amino-acids such as hydrophobicity, is in principle better 
applicable to proteins with a higher level of energetic frustration.
Moreover, requiring the knowledge of primary and secondary structures only, 
the model has a greater predictive power, and in this sense, 
could be considered closer to an ab initio representation. The above 
arguments motivate a detailed comparison between the these 
two protein models to assess their applicability 
and potentialities in the study of biomolecules.

\section{Theory and Methods}
When native state topology plays the relevant role in driving the folding 
process, many molecular details of protein structures can be mapped onto 
simplified coarse-grained models encoding the overall topology through the 
knowledge of the native contacts. These models, neglecting side 
chains and peptide groups, reduce a protein chain to its backbone, 
where aminoacids are assimilated to beads centred on their $\alpha$ carbon 
atoms. The G\={o} energy function, mimicking a perfect funnel landscape,  
assigns to the native state the lowest energy by simply promoting the formation 
of native interactions. Here we employ the force field proposed by 
Clementi \emph{et al.} 
\citep{Clementi} with distance cutoff $R_c=6.5$~\AA \citep{MJ} to
to identify native contacts in the structure 1NMV.pdb. 
A native contact means that the distance, $R_{ij}$, 
of C$_{\alpha}$ atoms 
relative to residues $i$ and $j$ ($|i-j|\ge 3$) is less than $R_c$ in the 
native state. This pair undergoes an attractive LJ-interaction 
\begin{equation}
V_{nat}(r_{ij}) = \epsilon \left[ 5 \left(\frac{R_{ij}}{r_{ij}} \right)^{12} -
6 \left (\frac{R_{ij}}{r_{ij}} \right)^{10} \right ]
\label{eq:Vnat}
\end{equation}
with equilibrium distance $R_{ij}$ . 
When two residues are not in a native contact ($R_{ij} > R_c$) they 
interact through a repulsive potential 
$$
V_{nnat}(r_{ij}) = \frac{2\epsilon}{3} 
\bigg( \frac{\sigma}{r_{ij}} \bigg)^{12}
$$ 
with $\sigma= 4.5$~\AA. 
These non-native interactions, besides ensuring the self-avoidance of the 
chain, generally enhance the cooperativity of the overall folding process 
\citep{Plotkin}. 
A further bias towards the native secondary structure is introduced through 
a bending and a dihedral potential. The former is modeled as an 
harmonic function and allows only small oscillations around the native 
angles ($\theta_{i}^0$) formed by three consecutive residues
$$
V_{\theta}(\theta_i) = \frac{1}{2}k_{\theta}(\theta_{i} - \theta_{i}^0)^{2}
$$
with harmonic constant $k_{\theta} = 20\epsilon$. 
The most important determinant of the secondary structure is the dihedral 
potential arising from the torsional energy. 
Each dihedral angle, identified by four consecutive beads, contributes to 
the potential with the terms 
$$
V_{\phi}(\phi_i) = k_{\phi}^{(1)}[1 - \cos(\phi_{i} - \phi_{i}^{0})] + 
k_{\phi}^{(3)}[1 - \cos3(\phi_{i} - \phi_{i}^{0})]
$$
where $\phi_{i}^{0}$ is the value of angle $i$ in the native structure, 
$k_{\phi}^{(1)} = \epsilon$ and  $k_{\phi}^{(3)} =0.5\epsilon$.
Finally, consecutive residues interact with each other
through the potential harmonic in their distance $r_{i,i+1}$ 
\begin{equation}
V_h(r_{i,i+1}) = \frac{k_h}{2}(r_{i,i+1} - b_i)^2 \, \, ,
\label{harmonic}
\end{equation}
which maintains the chain connectivity, with $b_i$ being the 
native bond-length     
and $k_h = 1000/r_0^2$.
Therefore, the global G\={o}-potential reads
\begin{eqnarray*}
V_{Tot} & = & \sum_{i=1}^{N-1}V_h(r_{i,i+1}) + 
\sum_{i=1}^{N-2}V_{\theta}(\theta_i) + \sum_{i=1}^{N-3}V_{\phi}(\phi_i) + \\
& + & \sum_{i,j>i+2}\{V_{nat}(r_{ij})\Delta_{ij} + 
(1 - \Delta_{ij})V_{nnat}(r_{ij})\}\;.
\end{eqnarray*}
where $\Delta_{ij} = 1 (0)$ if the contact is native (non-native).
G\={o} models of the type just outlined may produce a gradual folding  
behaviour being uncapable to reproduce the typical kinetic cooperativity 
of two-state folders. Experimental studies suggest \citep{Northey,DiNardo} 
that the origin of 
cooperativity lies in specific interactions 
appearing only after the assembly of native-like structures. 
These particular interactions can be modeled by imparting an extra energetic global 
stabilization to the native state \citep{Chan} through a different analytical 
form of the energy function when the chain visits the native basin. 
In the present work we need to implement the rescaling method to make
the folding transition highly cooperative in agreement with the experiments on the WW domain.
The interaction forces on the residues are thus computed according to the 
following rule
\begin{equation}
F_{conf} = \left\{ \begin{array}{ll}
                      -\nabla V_{Tot} & \mbox{for $Q < Q_{th}$} \\
                      -\nabla V_h - \rho\nabla(V_{Tot} - V_h) & 
\mbox{for $Q \ge Q_{th}$}                      
\end{array}
\right .
\end{equation} 
where $Q$ is the fraction of formed native contacts 
and $\rho = 2$ is the scaling factor.   
The force rescaling determines a higher free energy barrier between the 
folded and unfolded state in correspondence of the folding temperature 
resulting in a higher cooperativity. 
Therefore, the residence times in the folded and unfolded state are 
expected to be significantly longer. 

The SHG model is an off-lattice minimal model that generalizes a previous 
model introduced by Thirumalai and co-workers \citep{Thirumalai,Thirumalai2}. 
This approach represents $\alpha$-carbons with beads of three possible 
flavours hydrophobic (B), hydrophilic (L) and neutral (N), 
according to table \ref{tab:flavours}. 
\begin{table}[h]
\begin{center}
\begin{tabular}{|lc|lc|lc|lc|}
\hline
A.A.~ & F & A.A.~ & F & A.A.~ & F & A.A.~ & F \\
\hline
Ala &  B  & Met & B & Gly & N & Asn & L \\
Cys &  B  & Val & B & Ser & N & His & L \\
Leu &  B  & Trp & B & Thr & L & Gln & L \\
Ile &  B  & Tyr & B & Glu & L & Lys & L \\
Phe &  B  & Pro & B & Asp & L & Arg & L \\
\hline
\end{tabular}
\end{center}
\label{tab:flavours}
\caption{Dictionary for the translation
of three-letter code of the 20 natural amino-acid
into the three-flavor code \citep{SHG2}.}
\end{table}

The driving force responsible for the collapse onto a compact structure is the 
attraction between 
B-beads, whereas the repulsion between L and N beads determines the rearrangements of the 
compact structure into the native topology. 
The long-range interactions between residues which 
may be far apart in sequence space is modelled through the potential:
\begin{equation}
V_{LR} = \sum_{i,j\ge i+3} \epsilon_{h}
S_1 \left [ \left(\frac{\sigma}{r_{ij}}\right )^{12} -
2 S_2 \left(\frac{\sigma}{r_{ij}}\right )^{6} \right ]
\end{equation}
where $\epsilon_h$ ($1.65$ Kcal mol$^{-1}$ see below) sets the energy 
scale and $\sigma = 4.0$~\AA. 
The attractive forces between hydrophobic residues is attained by setting 
$S_1=S_2=1$ for BB pairs, while 
the interactions involving the LL and LB pairs are characterized by $S_1=1/3$ and $S_2=-1$. 
This interaction is repulsive and the $r^{-6}$ 
term, which accounts for the hydration shell around the hydrophilic residues, 
makes the potential 
longer ranged than the usual $r^{-12}$. 
The forces involving neutral residues are also repulsive and 
amount to an excluded volume potential by setting $S_1=0$ and $S_2=0$.
The secondary structure arises as a result of bending and dihedral 
interactions, which surrogate 
side-chain packing and hydrogen-bonding. The analytic expression of the dihedral potential is
\begin{eqnarray}
V_{dih} = &\sum_{i=1}^{N-3} [A_{i}(1 + \cos \phi_{i}) + B_{i}(1 - \cos \phi_i) + 
\nonumber 
\\ 
          & C_{i}(1 + \cos 3\phi_{i}) + D_{i}(1 + \cos(\phi_{i} + \pi/4))] 
\label{eq:Vphi}
\end{eqnarray}

where $\phi_i$ indicates the angle between the two adjacent planes identified by the 
positions of four consecutive beads. The information on secondary structures is 
systematically stored in the coefficients $A,B,C$ and $D$ that determine a bias on the 
angles reflecting the propensity of residues to form a specific secondary motif. 
Indeed, each dihedral, in the chain, is defined to be 
either helical (H: $A_i=0$, $B_i = C_i =D_i = 1.2\epsilon_h$), extended (E: $A_i=0.9 \epsilon_h$, 
$C_i=1.2\epsilon_h, B_i=D_i=0$), or turn (T: 
$A_i=B_i=D_i=0, C_i=0.2\epsilon_h$). Therefore, the primary structure must be complemented with 
the auxiliary sequence, of "E,H,T" symbols assigning the appropriate set of coefficients. 
The decoupling between primary and dihedral sequence, not present in similar models (40,41), increases the 
possibilities in the modulations of relative strengths between local and non-local interactions 
which results in a finer structural tuning \citep{SHG2}. 
The Head-Gordon force field is completed by a 
bond angle interaction modelled as a harmonic potential 
\begin{equation}
V_{\theta} = \sum_{i=1}^{N-2}\frac{k_{\theta}}{2}(\theta_i - \theta_0)^2
\end{equation}
with a constant $k_{\theta} = 20\epsilon_h/(rad)^2$ so that large deviations from 
the equilibrium value $\theta_0 = 1.8326$ rad are unlikely, and bond angles result 
basically fixed. Also in this model, stiff springs (\ref{harmonic}) with equilibrium 
distance $r_0 = 3.8$~\AA,  
maintain the chain connectivity mimicking the presence of covalent peptide bonds between 
successive aminoacids. 
This stiff interaction allows to keep the bond length approximately fixed, 
while being less computationally demanding than the RATTLE algorithm used in previous works 
\citep{Thirumalai,SHG3} to enforce fixed bond lengths.
The SHG model retains only the minimal number of elements needed to capture the essential 
features of protein molecules, however, some strong determinants such as hydrogen bonding and 
side chains are missing. These limitations should be compensated through a design strategy 
\citep{Design} 
for optimising the sequence. Here, we used the sequence LBBNN-BLBLB-NLNNN-LBBBB-
LLNNL-BNBBL-LBNNL proposed in Ref. \citep{SHG2} for the hPin1 WW domain and designed via a 
threading approach based on energy gap maximization (42). The secondary structure propensity, 
selecting the native-like dihedral angles, is encoded in the auxiliary sequence TTTTT-EEEEE-
TTTTT-TEEEE-TTTTT-EETTT-EET, built directly through the information contained in the 
PDB file 1NMB.pdb. 
In order to control the temperature, we performed constant temperature MD simulations within 
the isokinetic scheme \citep{Isokin} using dimensionless quantities. The temperature was 
measured in units of $\epsilon_h/R = 1070.96$ K, time in units of 
$t = \sigma(\epsilon_h/M)^{1/2}$= 4.44 ps, 
($\sigma= 4.0$~\AA~is the equilibrium length of  
Lennard-Jones interactions, $M=110$ is the average aminoacid mass), energy in 
units $\epsilon_h$, specific heat in units 
$R=1.9872\times10^{-3}$ Kcal mol$^{-1}$K$^{-1}$ and the radius of gyration in units $\sigma$. 
The energy scale $\epsilon_h$ was set to $1.65$ Kcal mol$^{-1}$ to reach a 
denaturation temperature compatible with experimental data \citep{Gruebele} 
$T=332$ K.
For the G\={o}-model the same units apply, except for the energy scale 
set to $\epsilon = 0.66$ Kcal mol$^{-1}$.  
During the simulations, we monitored the difference from the native state or 
reference state  through the overlap Q, representing the fraction of formed 
native contacts
$$
Q = \frac{\sum_{i,j\ge i+3} \Theta(R_c - r_{ij})}{\sum_{i,j\ge i+3}
                            \Theta(R_c -R_{ij})}
$$
where, the sum runs over all pairs of native contacts, $r_{ij}$ and $R_{ij}$ are the 
distances of residues $i$ and $j$ in the current and in the reference 
conformation respectively and $\Theta(u)$ indicates the unitary step function. 
A value $Q \cong 1$, 
indicates that the conformation is native-like, while values close to 
zero refer to denaturated states. 
we also considered, as further reaction coordinates of the folding/unfolding process, 
the gyration radius and the root mean square distance (RMSD) between the 
current and reference conformations after an optimal superposition performed according to 
Kabsch's algorithm \citep{Kabsch}.
The thermodynamics of the folding/unfolding transition was obtained via the 
weighted histogram method \citep{Ferren,Guo}. 
This technique offers the possibility to gain a better sampling of the 
conformation space than ordinary methods. The procedure consists in storing 
bidimensional histograms of the number of contacts $N(E,Q)$ as a function 
of the energy E and coordinate $Q$ at each temperature run. 
Such histograms are then optimally combined to reconstruct 
the best estimate of the density of states $\Omega(E,Q)$, which, in turn, 
will be used to compute the thermodynamics of the system. 
The knowledge of $\Omega(E,Q)$ can be also employed to derive the 
probability that, at temperature $T$, 
the protein states are characterized by energy $E$
and reaction coordinate $Q$
$$
P_T(E,Q) = \Omega(E,Q) \exp\{-\beta(E-F)\}
$$
where $\beta=1/RT$ and $F$ is the total free 
energy of the system coming from the normalization of  
$P_T(Q,E)$. 
The sum of $P_T(E,Q)$ over all possible energies E provides the 
probability for the system to have a specific value Q at 
temperature T, which in 
turn, by reversing the Boltzmann's weight, 
gives the potential of mean force along the reaction coordinate Q
$$
W_T(Q) = -RT \ln[P_T(Q)] \;.
$$
We computed the specific heat profile as a function of the temperature: 
its peaks and shoulders locate those temperatures at which the main 
structural chain rearrangements occur 
A detailed characterization of the folding/unfolding process can be 
obtained by measuring the probability of native contact formation as 
the temperature is varied 
$$
P_{ij}(T) = \langle \Theta(R_c -r_{ij})\rangle
$$
where the average $\langle \cdots \rangle$ is taken over time 
assuming the dynamics to be ergodic. $P_{ij}(T)$ typically features a 
sigmoidal shape, keeping values close to 1 at low 
temperatures and decreasing to zero at high temperatures. 
The knowledge of probabilities $P_{ij}(T)$  
allows for a classification and ranking of native contacts according to 
their "thermodynamic relevance" \citep{Cecconi,Cecconi2,Dokholyan} thus 
suggesting possible reaction pathway, key residues \citep{Vendruscolo} 
and folding nucleus \citep{Abkevich}.

\subsection{$\Phi$ values}
The comparison of G\={o} and the SHG models on the WW domain provide the 
opportunity to study the relevance
of topological versus energetic frustration \citep{Frustr} 
in the folding mechanism. This can be accomplished by  $\Phi$ values 
computation and by the further comparison with experimental data. 
$\Phi$-values \citep{Fersht} measure the 
perturbation effects of a site-directed mutation 
which, by altering the free energy difference among 
native, transition  and unfolded states may affect  
the thermodynamics and the kinetics of the reaction.
A prevalence of topological or energetic frustration may be argued from 
a better fit with the experiments of the 
G\={o}-derived or SHG-derived $\Phi$-values respectively 
\citep{Clementi,Vendruscolo}.    
The $\Phi$-values can be computed through a kinetic approach from the 
folding and unfolding rates of the mutant and wild-type protein 
\citep{FershtBook}:
\begin{equation}
\Phi =  \frac{RT\log(k_{f}^{WT}/k_{f}^{mut})}
{RT\log[(k_{f}^{WT}/k_{f}^{mut})\cdot(k_{u}^{mut}/k_{u}^{WT})]}
\label{kinet_phi}
\end{equation}
where $R$ is the ideal gas constant, $T$ is the absolute temperature and 
$k_{f}$ and $k_{u}$ are the folding and unfolding rates respectively. 
The denominator of the above expression is just the 
total stability change $\Delta\Delta G^{0}$. The use of 
equation~\ref{kinet_phi} is computationally demanding as it requires a 
simulation for each mutation. This motivates the use of a thermodynamic 
strategy for the $\Phi$-value evaluation \citep{FershtBook}:
\begin{equation}
\Phi = \frac{\Delta\Delta G^{\dagger}}{\Delta\Delta G^{0}} = 
\frac{\Delta\Delta G^{TS} -
\Delta\Delta G^{U}}{\Delta\Delta G^{F} - \Delta\Delta G^{U}}
\label{thermod_phi}
\end{equation}
where $\Delta\Delta G^{\dagger}$ is the change in stability of the 
free-energy barrier between the native and denaturated state. 
Equation~\ref{thermod_phi} is equivalent to equation~\ref{kinet_phi} 
when Kramer-like theory applies \citep{Frustr}.

If the effect of the mutations is sufficiently small, then, 
following Ref.~\citep{Clementi} the $\Phi$-values can be derived by 
a free energy perturbation (FEP) approach:
\begin{equation}
\Phi = 
\frac{
\log\langle\exp\{-\Delta E/RT \}\rangle_{TS} - 
\log\langle\exp\{-\Delta E/RT \}\rangle_{U}}
{\log\langle\exp\{-\Delta E/RT \}\rangle_{F} - 
\log\langle\exp\{-\Delta E/RT \}\rangle_{U}
}
\label{FEP_phi}
\end{equation}
where the Boltzmann factors depend on the energy difference between the 
mutant and the wild type (WT) and the averages are computed over 
WT-conformations of the folded (F), transition state (TS) and unfolded (U) 
ensembles.

In the present paper, the $\Phi$-values are computed according to 
equation~\ref{FEP_phi} using a method developed in Ref. \citep{Clementi} 
that can be summarized in the following steps:
\begin{enumerate}
\item Determination of the folding temperature $T_f$ from the specific 
heat plot.
\item Analysis of the free energy profile at temperature $T_f$ plotted as a 
function of a suitable reaction coordinate. 
The free energy profile of a two-state folder typically shows a double-well 
shape which allows to choose three windows of the reaction coordinate 
identifying the folded, transition state and unfolded ensembles respectively.
\item Dynamic simulation at $T = T_f$ and storage of conformations 
belonging to the F, TS and U ensembles.
\item Choice of mutations and computation of FEP $\Phi$-values.
\end{enumerate}
Structural information about the native-likeness of the transition state 
was also gained from the so-called structural $\Phi$-values 
\citep{Settanni}:
\begin{equation}
\Phi_{struc}(i) = 
\frac{1}{N_{j \in C(i)}}\frac{\sum_{j\in C(i)}\;P_{TS}(i,j)}
{\sum_{j\in C(i)}P_{F}(i,j)}
\label{eq:Phistruct}
\end{equation}
where $P_{F}(i,j)$ and $P_{TS}(i,j)$ are the frequencies of the native 
contact $i-j$ in the folded and 
transition ensembles respectively, and the sums run over the set 
$C(i)$ of native contacts in which residue $i$ is involved.

\section{Results}
We report on the thermodynamic properties and contact formation patterns 
observed in unfolding/refolding equilibrium MD simulations of 
the WW domain. We first analyze the simulations based on G\={o}-model and 
then we discuss the corresponding scenario in the SHG-model approach.
Since the implementation of the SHG model requires a well designed
sequence we employed the 6-40 truncated sequence already optimized
in Ref.~\citep{SHG2}. For the sake of a consistent comparison 
with G\={o}-simulations, 
the corresponding fragment was extracted from the NMR structure 
stored in the PDB file 1NMV~\citep{Bayer}. 

\subsection*{G\={o}-model} 
\begin{figure}[!h]
   \begin{center}
      \includegraphics*[width=\columnwidth]{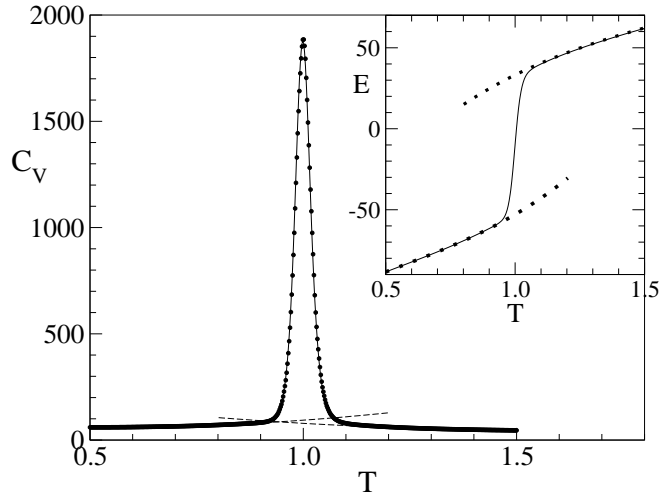}
   \end{center}
\caption{Heat capacity as a function of temperature in G\={o}-model
              simulations: folding (solid), unfolding (dotted). 
              Inset: thermal behaviour of energy; dotted lines represent
              quadratic fits of the baselines.}
\label{fig:go_Cv}
\end{figure}

A folding simulation was performed through a gradual cooling of a random
coil structure from a temperature $T=1.5$ down to $0.5$ in forty steps.
The specific heat profile, Fig.~\ref{fig:go_Cv}, is characterized by a single 
narrow peak at temperature $T_f=1.0$ suggesting a possible two state process. 

The same conclusion can be drawn from the ratio of
van't Hoff over the calorimetric enthalpy changes 
amounting to 0.74 without and $0.99$ with standard baseline subtraction 
\citep{baseline}.
\begin{figure}[!h]
   \begin{center}
      \includegraphics*[width=\columnwidth]{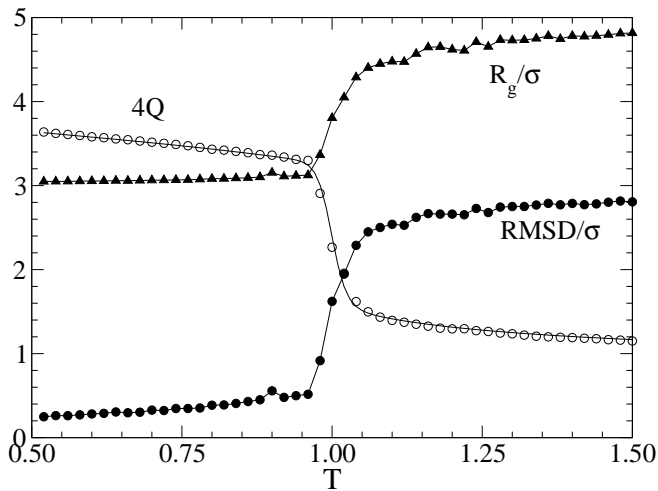}
   \end{center}
\caption{Structural parameters monitored during the G\={o}-model
         folding simulations. Triangles are the reduced gyration radius;
         filled circles indicate reduced $RMSD$; open circles refer to 
         the fraction of native contacts $Q$ magnified by a factor $4$. 
         Each point in the plots corresponds to an average of 
         $6\times10^{6}$ conformations sampled every $10^{3}$ time-steps.
         RMSD and $Q$ are computed using the PDB structure as a 
         reference.}
\label{fig:go_KQR}
\end{figure}

\begin{figure}[!h]
   \begin{center}
      \includegraphics*[width=\columnwidth]{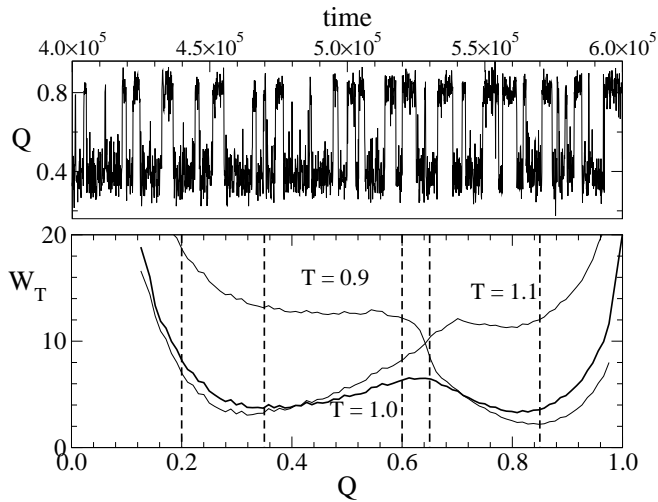}
   \end{center}
 \caption{Free energy profiles at different temperatures as a function of
         the overlap $Q$ (Lower panel). Vertical lines indicate the 
         the boundaries of the sampling windows for F ($ 0.85<Q<1.00 $)  
         U ($0.20 <Q< 0.35 $) and TS ($0.60 <Q<0.65 $) see text.
         Upper panel: time evolution of $Q$ at the folding temperature
         $T=1$ oscillating between the minima of the free energy wells.}  
\label{fig:go_jump}
\end{figure}

The folding/unfolding processes are reversible in temperature as shown 
by the agreement between specific heat plots.
The other observables used to characterize the folding transition such as, 
RMSD, overlap and gyration radius exhibit an abrupt change in correspondence 
to the folding temperature $T_f$ (Fig.~\ref{fig:go_KQR}). 
Free energy profile (Fig~\ref{fig:go_jump}) as a function of the overlap,
around the folding temperature, clearly features two distinct wells 
identifying the folded and unfolded ensembles separated by a 
barrier corresponding to the transition state conformations. 

\begin{figure}[!h]
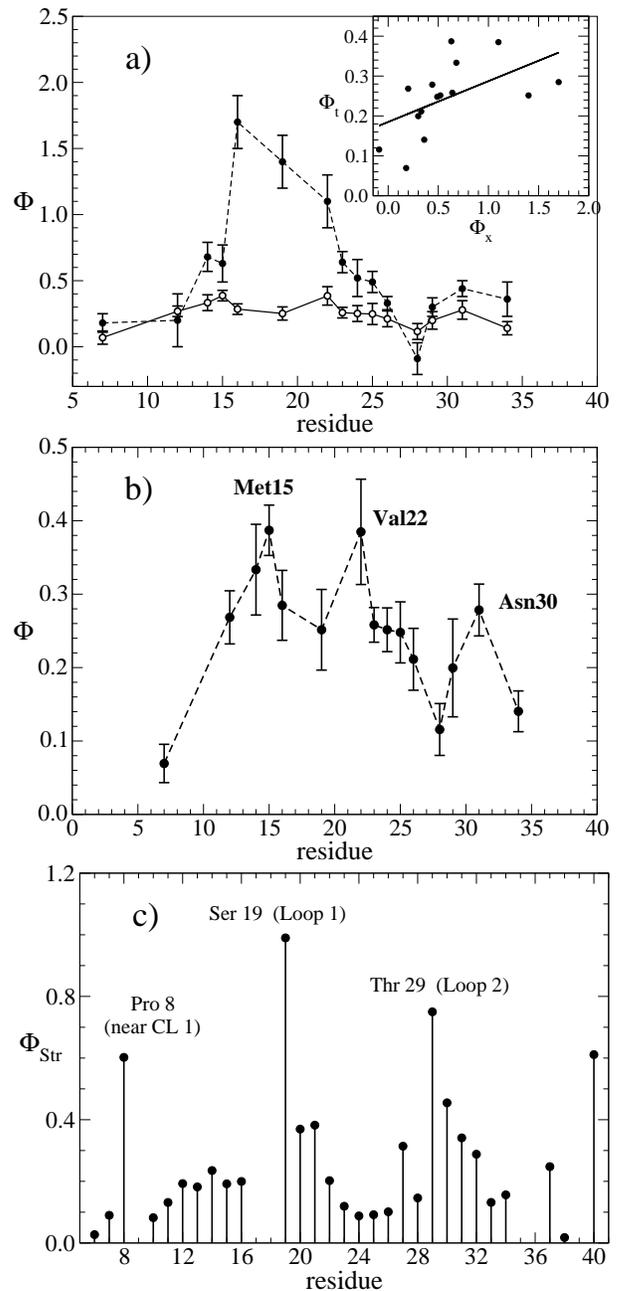

   \begin{center}
      \includegraphics*[width=0.92\columnwidth]{Fig5a.eps}
      \includegraphics*[width=0.92\columnwidth]{Fig5b.eps}
      \includegraphics*[width=0.92\columnwidth]{Fig5c.eps}
   \end{center}
\caption{a) Comparison between experimental (solid circles) and 
         theoretical (open circles) $\Phi$-values restricted to the 
         mutations performed in Ref.~\citep{Gruebele} except for Ser18. 
         $\Phi$-values were computed from the
         conformations sampled in a G\={o} simulation at folding 
         temperature using the perturbation method (\ref{thermod_phi}). 
         The inset shows the linear regression analysis
         between the two data sets with a correlation coefficient $0.54$.
         b) Enlargement of theoretical $\Phi$-value shown in panel a). 
         c) Structural $\Phi$-values computed from formula 
         (\ref{eq:Phistruct}) using the (F) and (TS) ensemble 
         structures.
         The three peaks in the plot indicate that the two loops and the 
         first hydrophobic cluster are native-like in the transition state.}
\label{fig:go_phival}
\end{figure}

The shape of the free-energy plot suggests
a choice of overlap windows for the sampling of conformations in the 
three ensembles F, U and TS (see caption of Fig.\ref{fig:go_jump}) 
for the computation of $\Phi$-values (Methods).
In figure~\ref{fig:go_phival} we compare our single site simulated $\Phi$-values
(Eq.~\ref{FEP_phi}) with the experimental data by Gruebele \citep{Gruebele}. 
In the G\={o}-like approach, a mutation can be modeled as the removal of a 
single native contact \citep{Clementi} or in alternative, as an average over 
all possible removals of contacts involving the same residue. We followed the
second strategy considering only contacts $|i-j|\ge 3$.  In this scheme 
we cannot evaluate $\Phi$-value of Ser18 because it lacks such contacts.
The theoretical $\Phi$-values in figure ~\ref{fig:go_phival} 
vary in the range $[0.0,0.5]$, whereas the experimental ones are distributed in
a much wider interval. This feature is an expected result of the very limited
energetic frustration of the G\={o}-force fields \citep{Frustr}. 

The discrepancy is reflected by 
the modest value of the linear correlation coefficient $r = 0.54$, (see 
regression line in the figure~\ref{fig:go_phival}a inset). 
Of course we cannot exclude that a possible improvement of $\Phi$-value 
accuracy might be achieved either by 
employing other mutation implementations or by using   
alternative contact
maps accounting for the high flexibility of the native structure of peptides 
and small proteins \citep{Cavalli}.
Despite this  
not high correlation, the 
theoretical $\Phi$-values provide a qualitative indication about the molecule 
regions that are still native-like in the transition state. 
The plot in fact indicates that the sites most sensitive to mutations are those 
in the region of loops L1 and L2 in agreement with experimental 
results (see Figure~\ref{fig:go_phival}b).

The picture provided by the structural $\Phi$-values 
(Fig.~\ref{fig:go_phival}c) 
is consistent with that derived from the perturbation method, in
fact also in this case the highest $\Phi$-values correspond to residues 
located in L1 (Ser 19), L2 (Thr 29) or in the neighborhood of the first 
hydrophobic cluster CL1 (Pro 8). The low $\Phi$-values pertain mainly
to residues in strands $\beta_1$ and $\beta_2$ suggesting that these two 
regions are unlikely to be in contact in the transition state.

\subsection{SHG-model}

Ten independent folding simulations starting from random-coil conformations, 
were performed through a gradual cooling schedule from temperature 
$T =1.0$ to $T = 0.01$ in $40$ steps. 
The final structures were further relaxed
by a steepest descent cycle until the maximal total force per monomer 
reached a value smaller than $10^{-8}$ Kcal mol$^{-1}$\AA$^{-1}$. 
We obtained different folds and chose the conformation with lowest 
energy ($E=-19.0035\epsilon$) and lowest RMSD (4.74~\AA)  
from the PDB structure as the reference structure $\Gamma_0$ 
(Fig.~\ref{fig:struct}). 
However, the simulations revealed also the 
existence of another degenerate minimum with the same energy and 
specular to $\Gamma_0$ resulting in much higher RMSD.

Despite the large value of RMSD, $\Gamma_0$ correctly displays the 
topology of a triple-stranded, antiparallel $\beta$-sheet, that however 
lacks the typical twist of the PDB structure making 
loop L2 almost perpendicular to loop L1 (see Figure \ref{fig:struct}). 
As a result, the folded structure is much more compact than the real 
protein and has a much larger number of native contacts (71 versus 41). 
The fact that 22 out of the 41 PDB 
contacts are also present in the folded structure is an indication 
of the satisfactory structural performance of the SHG simulation.
 
Structure $\Gamma_0$ was then denaturated through ten independent 
runs with the same but inverse temperature schedule, involving a 
thermalization stage of $6 \times 10^{6}$ time steps ($\Delta t=0.005$) 
at each temperature followed by a run over the same length, where control 
parameters were measured to assess the unfolding progress.
The course of both folding and unfolding simulations was monitored through 
the analysis of the energy, the specific heat, RMSD from $\Gamma_0$, the 
overlap and radius of gyration $R_g$.

\begin{figure}[!h]
   \begin{center}
      \includegraphics*[width=\columnwidth]{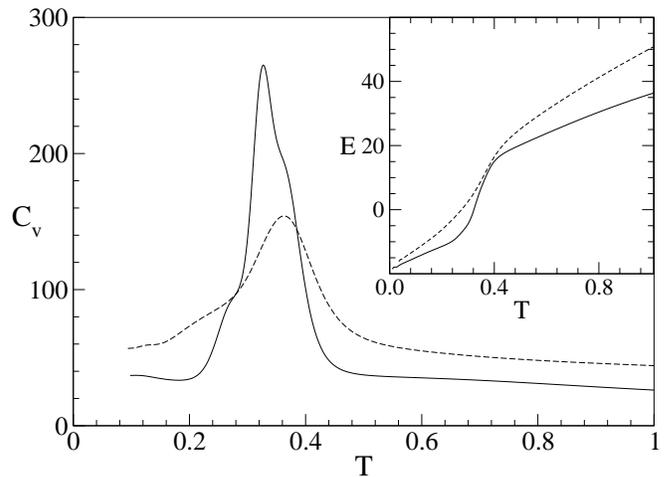}
   \end{center}
\caption{Thermal behaviour of heat capacity (main figure) 
         and energy (inset) during unfolding (solid lines) 
         and folding (dashed lines) SHG simulations. The 
         thermodynamic observables have been computed using
         the weighted histogram method.}
\label{fig:shg_cv}
\end{figure}

Both the folding and unfolding specific heat plots (Fig.~\ref{fig:shg_cv}) are characterized by the 
presence of a main peak and a shoulder. The peaks of the folding and unfolding
thermograms $P_f$ and $P_u$ are located at $T_f = 0.36$ and $T_u = 0.33$
respectively whereas the shoulders $S_f$ and $S_u$ correspond to $T_{Sf} = 0.24$
and $T_{Su} = 0.28$.  The folding process appears not to be fully reversible 
probably due to the fact that the sequence, although designed, 
is not yet a good folder. 

\begin{figure}[!h]
   \begin{center}
      \includegraphics*[width=\columnwidth]{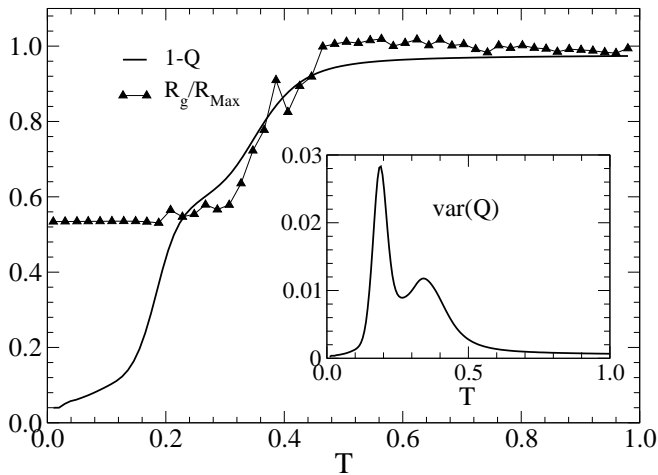}
   \end{center}
\caption{Noncooperativity of the SHG folding simulation yielding our 
         best final structure $\Gamma_0$. After a first collapse the
         radius of gyration remains constant whereas the structural
         difference from the reference structure $1-Q$ keeps on 
         decreasing at temperatures corresponding to the shoulder of the
         $C_V$ plot signalling a massive structural rearrangement. 
         Inset: temperature dependence of fluctuations of the structural 
         overlap. The main peak, located at the same temperature as the 
         heat capacity shoulder, corresponds to the folding temperature.}
\label{fig:shg_two_steps}
\end{figure}

The existence of the shoulder in the folding $C_v$ plot is a signature of a 
non cooperative folding mechanism in which, an initial collapse is 
followed by a structural chain rearrangement characterized by a significant 
increase in the number of native contacts 
unaffecting the overall  compactness of the molecule 
(see Fig.~\ref{fig:shg_two_steps}).  

This is confirmed by the thermal fluctuation of the structural overlap 
$$
\mbox{var}_Q(T) = \langle Q^2 \rangle - \langle Q \rangle^2 
$$ 
featuring the highest peak, not in correspondence of the main peak $P_f$ but
at the temperature $T_{Sf}$ of the shoulder (inset of 
Fig.~\ref{fig:shg_two_steps}). 

\begin{figure}[!h]
   \begin{center}
      \includegraphics*[width=\columnwidth]{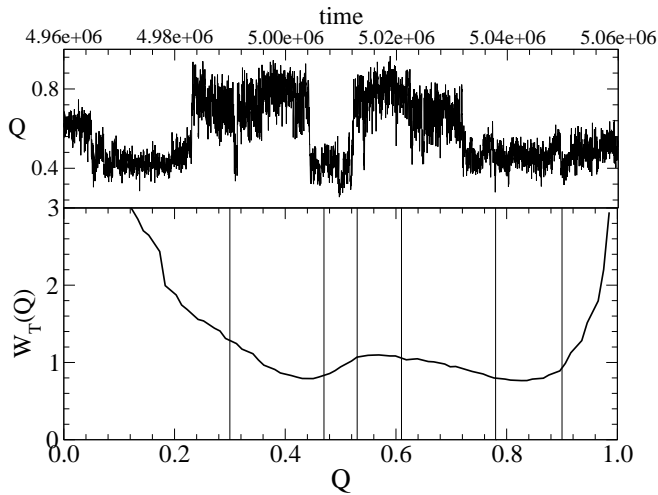}
   \end{center}
\caption{Free energy profile as a function of the overlap $Q$ at the 
         folding temperature $T = 0.237$  
         (lower panel). The vertical lines mark the boundaries of the 
         Folded ($0.78<Q<0.90$), Transition State ($0.53<Q<0.61$) and
         Unfolded ($0.30<Q<0.47$) ensembles. The upper panel shows the
         typical temporal evolution of the overlap in a sub-interval 
         of a simulation at folding temperature. The overlap was sampled
         every $5\times10^{3}$ time-steps.}
\label{fig:shg_jump}
\end{figure}

\begin{figure}[!h]
   \begin{center}
      \includegraphics*[width=\columnwidth]{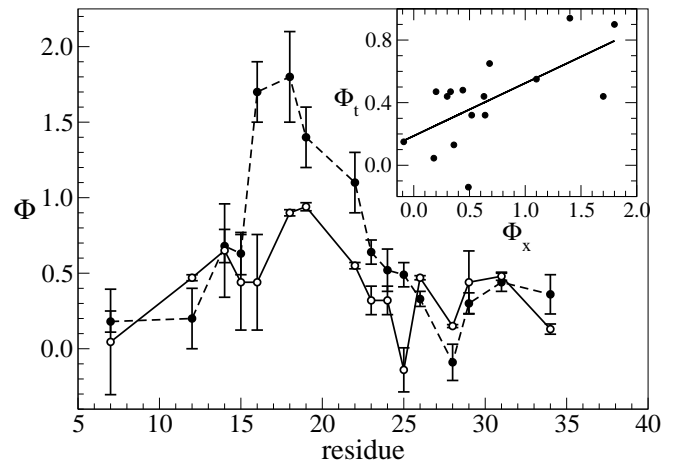}
   \end{center}
\caption{Experimental (solid circles) and computed (open circles)
         $\Phi$-values. $\Phi$-values were computed from the
         conformations sampled in a SHG simulation at folding
         temperature using the perturbation method. The two profiles
         show a qualitative agreement although the correlation
         coefficient of the regression line (see inset) is 
         $r = 0.65$.}
\label{fig:shg_phi}
\end{figure}

The marked difference between the folding and unfolding specific heat 
suggests the opportunity to consider the free energy 
profiles $W_T(Q)$ to better determine the folding temperature. 
The profiles (lower panel of Fig.~\ref{fig:shg_jump}) 

indicate that the transition is characterized by the presence of two wells 
separated by a barrier and the temperature where these two wells are evenly 
populated is $T=0.237$. This confirms that, the peak of $C_v$ is mainly 
related to the $\Theta$-collapse, whereas the shoulder corresponds to the 
folding transition. 
In fact, kinetic simulations at temperature $T=0.237$ show that the time 
evolution of $Q(t)$ exhibit jumps between the two 
free energy wells (upper panel of Fig.\ref{fig:shg_jump}). 
The double-well shape of the free-energy profile again allows to sample 
conformations in the {\em folded} (F), {\em transition state} (TS)  
and {\em unfolded} (U) ensemble used to implement the perturbation technique 
for $\Phi$-value computation (Methods). 
The plot  
in figure \ref{fig:shg_phi} shows the $\Phi$-values restricted to the set of 
residues mutated by Gruebele \citep{Gruebele}. 

For each site we tested the effect of all the possible single mutations 
allowed by the model, namely two hydrophobicity  shifts and two shifts in the 
secondary structural bias for each of the two dihedral angles flanking the 
residue under examination. For each site we chose the least perturbative 
mutations.
Theoretical $\Phi$-values feature two major peaks in correspondence with 
loop L1 and loop L2 which is a qualitative resemblance with experiments. 
A more quantitative comparison is provided by the correlation coefficient 
between theoretical and experimental data amounting to $r = 0.65$.

\begin{figure}[!h]
   \begin{center}
      \includegraphics*[width=\columnwidth]{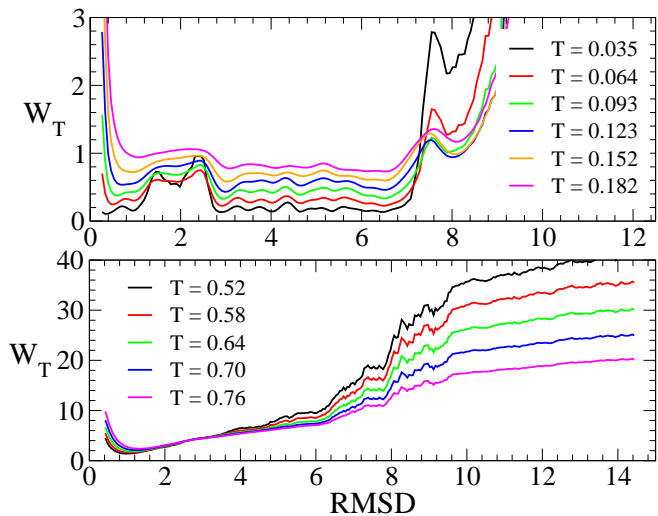}
   \end{center}
\caption{Upper panel: low-temperature free energy profiles of
        the SHG model as a function of the RMSD from the reference
          conformation $\Gamma_0$. The native valley appears to be
          partitioned in two main sub-basins separated by a barrier.
          The sub-basin corresponding to the RMSD range $[0.25-1.00]$
          is populated by conformations with the same chirality as the
          PDB structure, whereas the sub-valley in the range
          $[2.80-6.50]$, corresponds to the opposite chirality.
          Lower panel: low-temperature free energy profiles of the
          G\={o} model as a function of the RMSD from the
         native conformation (pdb-id = 1NMV). The native
         valley shows a single basin as opposed to the
         partitioning in two sub-valleys typical of the
         G\={o} model. }
\label{fig:shg_go_freeK}
\end{figure}

The set of native conformations collected during the kinetic simulation 
provides a structural characterization of the ensemble F whose  
most interesting feature is the clustering of native-basin conformations  
in two main subsets characterized by non overlapping distributions of RMSD  
from the reference structure $\Gamma_0$. 
This is a further indication of the high level of frustration of the 
free-energy landscape associated to the sequence and it is in agreement with
the findings by Miller and Wales \citep{Wales} about the glass-like 
structure of the energy landscape in a closely related model 
\citep{Thirumalai2}.  
This partitioning of the native basin is evident in 
figure~\ref{fig:shg_go_freeK}, 

where we plot the free energy versus the RMSD which is a structural indicator 
more sensitive than the overlap. A finer analysis reveals that the structures
in the two sub-basins of the native valley correspond to different  
chiralities but similar energies. 
The absence of such a partitioning in the same plot for the G\={o} force field,
indicates that this feature is mainly peculiar of the model rather than this
specific protein. 
 
To clarify how the landscape properties affect the reversibility
of the folding process we studied the folding/unfolding transition 
from the contact formation probabilities $P_{ij}$.        
In particular, as the plots $P_{ij}(T)$ are typically sigmoid, 
a contact can be regarded as broken/formed in correspondence of
the temperature, where the absolute value of the slope of the 
probability curve is maximal. This
allows to identify the contacts whose formation/breakdown occurs at a given 
temperature. 

\begin{figure}[!h]
   \begin{center}
      \includegraphics*[width=\columnwidth]{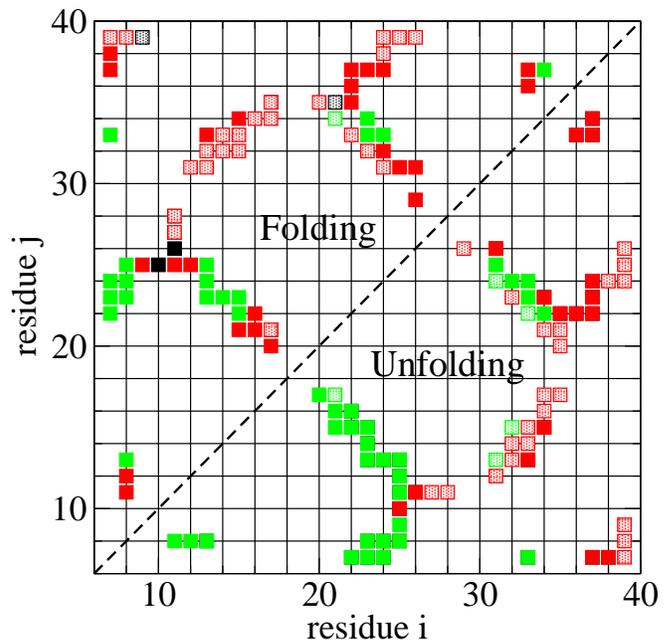}
   \end{center}
\caption{Contact maps summarizing the folding and unfolding SHG 
         process. The color code identifies three temperature ranges:
         black, $T < 0.15$; red, $0.15 \le T \le 0.30$; green, 
         $T > 0.30$. Shaded symbols refer to weak contacts with low 
         probability of formation ($P_{ij} < 0.5$) at the lowest 
         simulation  temperature $T = 0.01$.}
\label{fig:shg_path}
\end{figure}

To analyze the folding/unfolding process, we considered three 
temperature windows corresponding to different regions of the specific heat 
plots. The first window ($T < 0.15$) refers to the pre-transition baseline 
of the $C_v$ plot, the second window ($0.15 \le T \le 0.30$) insists on the 
region of the shoulder, and the third window ($T > 0.30$) includes the main 
peak. 
The contacts appearing or disappearing in correspondence of the three 
windows are shown in black, red and green respectively, in the contact maps
(Fig.~\ref {fig:shg_path})  
summarizing the main events of the pathway. Shaded symbols represent weak
interactions with probability of formation below  $50 \%$ at the lowest
simulation temperature $T = 0.01$.

The contact map shows that the first contacts formed during folding are 
located in the intermediate part of sheet $\beta_1$-$\beta_2$ and in the 
region of sheet $\beta_2$-$\beta_3$ most distant from loop 2. 
The formation of these contacts is responsible for the collapse of the molecule 
into a compact but not completely folded conformation. 
The map also shows that the shoulder of the $C_v$ plot is characterized 
by the zipping of sheets $\beta_1$-$\beta_2$ and $\beta_2$-$\beta_3$ towards 
loop 1 and loop 2 respectively. During the process there also 
occurs the locking of $\beta_1$-$\beta_3$, $\beta_2$-$tail$ and 
$head$-$\beta_2$ contacts (hereafter, the term 'head' and 'tail' indicate 
the amino-terminal region $Lys6-Gly10$ and the carboxy-terminal residues $Asn36-Asn40$
respectively).     
These contacts are not present in the 
PDB structure and they arise as a consequence of the higher compactness
of $\Gamma_0$, so that their formation probabilities are always below
$50 \%$. The folding is completed by the appearence of a few contacts between 
residues very far from each other along the protein chain.

During the unfolding reaction no "native" (with respect to $\Gamma_0$) 
contact breaks down in the low temperature window because the heating schedule 
enables the protein to escape easily from kinetic traps making the  
process much less gradual than folding. This reflects on  
the smaller number of contacts broken in the shoulder region as compared to 
the number of contacts formed in the same temperature range during folding. 
In particular, the cleavage occurs of 
$\beta_1$-$\beta_3$, $\beta_2$-$tail$ and $head$-$\beta_2$ contacts, 
whereas the dissolution of the contacts of loop 1 and loop 2 is delayed to 
the region of the peak of the $C_v$ plot 
where most $\beta_1$-$\beta_2$ and $\beta_2$-$\beta_3$ contacts also 
disappear.

The comparison of the two contact maps thus reveals that the sequences of 
the molecular events in the folding and unfolding processes are basically 
reverse to each other even if the unfolding is a more abrupt phenomenon 
occurring in a narrower temperature window.

\section{Discussion}
Our results indicate how the different approaches of structure-based and 
sequence-based description exemplified by  G\={o} and SHG model 
are appropriate to simulate complementary features of the folding
process.    
The G\={o} model, in fact, being based on the influence of the native state 
topology on the folding process is independent from the amino-acid 
sequence and it completely disregards the chemical properties of the molecule. 
The SHG model, on the other hand, is a minimal model, where the chemical 
features of amino-acids are partially included determining the folding 
driving force. 

Our simulations showed that the G\={o} model with angular bias \citep{Clementi} 
and rescaling \citep{Chan} can correctly reproduce the reversible, cooperative, 
two-state~mechanism of folding of hPin1 WW domain \citep{Gruebele}. 
The reversibility, indeed appears from 
the almost perfect superposition of the $C_V$ plots of folding and unfolding. 
Several elements, on the other hand, suggest a cooperative, two-state 
mechanism: 
the $C_V$ plots show a single sharp peak, the ratio of the van't Hoff to 
calorimetric enthalpy is close to 1 ($\kappa_{2}^{(s)} = 0.99$) \citep{ChanK2},   
all the indicators used to monitor the similarity with the native state 
exhibit a sharp sigmoidal thermal behaviour and the barrier 
between the two free-energy wells at the folding temperature is very high 
\citep{Coop,Coop2}.
The results from the simulation using the SHG model were rather
ambiguous. The simulated thermograms featured not only a peak, but
also a shoulder at lower temperature. This is the signature of a non 
cooperative folding involving a collapse into a compact, only partially 
structured globule, followed by a rearrangement into a native 
conformation. This scenario is confirmed by the thermal behaviour of the 
structural parameters (overlap, gyration radius and RMSD from native structure)
used to monitor the folding reaction. 
The results are consistent with the findings by Nymeyer 
\emph{et al.}~\citep{Critica} and by Guo and Brooks III~\citep{Guo}, in their 
simulations on the model by Honeycutt and Thirumalai \citep{Thirumalai}. 
The SHG formulation, although being an improvement of the latter model, 
still retains some of its drawbacks. Indeed, the conformations of the native 
ensemble (F), sampled at the folding temperature, can be clustered in two 
groups with non-overlapping RMSD distributions and opposite chiralities 
\citep{Cieplack}. 
The existence of two distinct clusters of native-like conformations 
can be easily explained by examining the low-temperature free-energy 
profiles as a function of the RMSD (from reference structure $\Gamma_0$): 
the native basin 
appears to be partitioned in sub-basins separated 
by barriers. 
The partitioning of the native basin is likely a feature that the SHG 
model inherited from Thirumalai model. Miller and Wales~\citep{Wales}, 
in fact, analyzed the disconnectivity graph of the potential energy 
surface of Thirumalai's force field drawing the conclusion that 
the energy hypersurface is not a single funnel, but it contains low-energy 
minima separated by high barriers.

Presumably, the reason for the degeneration of the native state of the 
SHG model 
relies in the symmetry of the dihedral potential $V_{\phi}$, Eq.
\ref{eq:Vphi}.  In particular, the sequence designed to represent the 
hPin1 WW domain, contains only "Extended" or "Turn" symbols, so that
$V_{\phi}$ is a polynomial in $\cos(\phi)$ and becomes symmetric for
the inversion $\phi \to -\phi$. 
The symmetry of the $V_{\phi}$ term, however, is not the only reason for 
the poor performance of the SHG model. In fact, we find that the energy 
histograms of the Folded- and Unfolded- state ensembles are significantly 
overlapping thus suggesting the existence of many low-energy, 
non-native conformations \citep{Guo}. 

We suspect that this is an effect of the only approximated
maximization of the energy gap between the native conformation
and the decoy set used in the sequence optimization procedure \citep{SHG2}. 
This would call for further refinements of the threading procedure.

Despite the several drawbacks, the SHG model enabled the 
computation of perturbation $\Phi$-values in qualitative agreement 
with experimental data. The linear correlation 
coefficient between theoretical and experimental $\Phi$-values ($r = 0.65$) 
is actually better than the one yielded by the G\={o} simulation ($r = 0.54$). 
The explanation of these result must be sought in the partial 
incorporation of the chemistry in the SHG description. Indeed, 
real mutations are chemical transformations of the molecule and  
they are better simulated by a chemically-based model such as the SHG rather 
than by a topological model.   
In the G\={o} model, in fact, mutations are generally simulated by 
the removal of native contacts \citep{Clementi}, however, they may affect 
all the interactions a residue is involved in. 
The SHG model, conversely, offers the possibility 
to treat mutations in a more realistic way because it implements 
shifts in the hydrophobic character of residues or changes
in the secondary structural bias of dihedral angles.
Moreover, the better agreement of experimental data with the SHG-computed 
$\Phi$-values may show that the folding mechanism of hPin1 WW domain is 
controlled not only by topological but also by energetic factors.

The significant differences, beyond statistical errors,  
 between the $\Phi$-values profiles
 yielded by the two models, in our opinion, reflect the different 
 strategies upon which the two models are built. 

A final issue that deserves some discussion is the quality of the 
structural prediction using the SHG
model. The SHG is a minimal model based on 
chemical properties of the system and a good outcome of the simulation 
is not \emph{a priori} 
guaranteed. The simulations show that, apart from chirality problems, 
the best final structure
$\Gamma_0$ (Fig.~\ref{fig:struct}) presents the correct topology of a 
three-stranded antiparallel 
$\beta$-sheet of the hPin1 WW domain, even if the structure appears to be 
more compact. 
This however, does not prevent the correct formation of 
both hydrophobic clusters. Moreover, $\Gamma_0$ shares 22 of the 41 
native contacts of the PDB 
structure.  

\section{Conclusions}

We performed folding and unfolding simulations of the WW domain of hPin1 
protein that represents an excellent candidate to test folding algorithms 
and models due to the availability of a large amount of structural, 
thermodynamical and kinetic experimental data.
The purpose of the work was to compare the performance of 
the G\={o} and SHG models that represent two different strategies to the 
folding problem.
Our simulations indicated that for the specific WW domain considered in 
this work, 
the G\={o} model with angular bias and 
rescaling, correctly reproduces the cooperative, two-state, reversible 
folding mechanism, whereas the SHG model does not. 
The reasons for the limitations of the SHG model must be sought in the 
insufficient optimization the sequence and in the non-funnel shape of the 
landscape. 
As a consequence, the present version of the SHG model does not allow reliable 
predictions of the folding mechanism. 
The satisfactory performance of the SHG model in the 
computation of $\Phi$-values, however, clearly shows the importance of 
incorporating the chemical properties of the sequence in a protein model. 
Our work, highlighting the limits of the SHG model, is thus intended to be 
a starting point for a further refinement of the model, in the firm belief 
that coarse-grained, minimal models represent viable alternatives
to computationally demanding all-atom simulations  
in investigations of large-sized, slow-folding proteins. 




\begin{thebibliography}{57}
\expandafter\ifx\csname natexlab\endcsname\relax\def\natexlab#1{#1}\fi

\bibitem[{Ferguson et~al.(2001)Ferguson, Johnson, Macias, Oschkinat, and
  Fersht}]{Ferguson}
Ferguson, N., C.~M. Johnson, M.~Macias, H.~Oschkinat, and A.~R. Fersht. (2001).
\newblock Ultrafast folding of ww domains without structured aromatic clusters
  in the denatured state.
\newblock \emph{Proc. Natl. Acad. Sci. USA} 98:13002--13007.

\bibitem[{Karanicolas and Brooks-III(2003)}]{BrooksIII}
Karanicolas, J., and C.~L. Brooks-III. (2003).
\newblock Structural basis for biphasic kinetics in the folding of the ww
  domain from a formin-binding protein: Lessons for protein design?
\newblock \emph{Proc. Natl. Acad. Sci. USA} 100:3954--3959.

\bibitem[{Verdecia et~al.(2000)Verdecia, Bowmanm, Lu, Hunter, and
  Noel}]{Verdecia}
Verdecia, M.~A., M.~E. Bowmanm, K.~P. Lu, T.~Hunter, and J.~P. Noel. (2000).
\newblock Structural basis for phosphoserine-proline recognition by ww domains.
\newblock \emph{Nature Struct. Biol.} 7:639--643.

\bibitem[{Bayer et~al.(2003)Bayer, Goettsch, Mueller, Griewel, Guiberman, Mayr,
  and Bayer}]{Bayer}
Bayer, E., S.~Goettsch, J.~W. Mueller, B.~Griewel, E.~Guiberman, L.~M. Mayr,
  and P.~Bayer. (2003).
\newblock Structural analysis of the mitotic regulator hpin1 in solution:
  Insights into domain architecture and substrate binding.
\newblock \emph{J. Biol. Chem} 278:26183--93.

\bibitem[{J\"ager et~al.(2001)J\"ager, Nguyen, Crane, Kelly, and
  Gruebele}]{Gruebele}
J\"ager, M., H.~Nguyen, J.~C. Crane, J.~W. Kelly, and M.~Gruebele. (2001).
\newblock The folding mechanism of a beta-sheet: the ww domain.
\newblock \emph{J. Mol. Biol.} 311:373--393.

\bibitem[{Bruscolini and Cecconi(2005)}]{Bruscolini}
Bruscolini, P., and F.~Cecconi. (2005).
\newblock Analysis of pin1 ww domain through a simple statistical mechanics
  model.
\newblock \emph{Biophys. Chem.} 115:153--158.

\bibitem[{Garnier et~al.(1996)Garnier, Wills, Verderame, and
  Sudol}]{Pin1-pathol1}
Garnier, L., J.~W. Wills, M.~F. Verderame, and M.~Sudol. (1996).
\newblock Ww domains and retrovirus budding.
\newblock \emph{Nature} 381:744--745.

\bibitem[{Sudol(1996)}]{Pin1-pathol2}
Sudol, M. (1996).
\newblock Structure and function of the ww domain.
\newblock \emph{Prog. Biophys. Mol. Biol.} 65:113--132.

\bibitem[{G\={o} and Scheraga(1976)}]{Gomodel}
G\={o}, N., and H.~A. Scheraga. (1976).
\newblock On the use of classical statistical mechanics in the treatment of
  polymer chain conformations.
\newblock \emph{Macromolecules} 9:535--542.

\bibitem[{Sorenson and Head-Gordon(2000)}]{SHG1}
Sorenson, J.~M., and T.~Head-Gordon. (2000).
\newblock Matching simulation and experiment: a new simplified model for
  simulating protein folding.
\newblock \emph{J. Comp. Bio.} 7:469--481.

\bibitem[{Brown et~al.(2003)Brown, Fawzi, and Head-Gordon}]{SHG2}
Brown, S., N.~J. Fawzi, and T.~Head-Gordon. (2003).
\newblock Coarse-grained sequences for protein folding and design.
\newblock \emph{Proc. Natl. Acad. Sci. USA} 100:10712--10717.

\bibitem[{Micheletti et~al.(1999)Micheletti, Banavar, Maritan, and Seno}]{Mich}
Micheletti, C., J.~R. Banavar, A.~Maritan, and F.~Seno. (1999).
\newblock Protein structures and optimal folding from a geometrical variational
  principle.
\newblock \emph{Phys. Rev. Lett.} 82:3372--3375.

\bibitem[{Mun\~{o}z et~al.(1998)Mun\~{o}z, Henry, Hofrichter, and
  Eaton}]{Eaton}
Mun\~{o}z, V., E.~R. Henry, J.~Hofrichter, and W.~A. Eaton. (1998).
\newblock A statistical mechanical model for beta-hairpin kinetics.
\newblock \emph{Proc. Natl. Acad. Sci. USA} 95:5872--5879.

\bibitem[{Alm and Baker(1999)}]{Baker}
Alm, E., and D.~Baker. (1999).
\newblock Prediction of protein-folding mechanisms from free- energy landscapes
  derived from native structures.
\newblock \emph{Proc. Natl. Acad. Sci. USA} 96:11305--11310.

\bibitem[{Fersht(2000)}]{Fersht}
Fersht, A.~R. (2000).
\newblock Transition-state structure as a unifying basis in protein-folding
  mechanisms: contact order, chain topology, stability, and the extended
  nucleus mechanism.
\newblock \emph{Proc. Natl. Acad. Sci. USA} 97:1525--1529.

\bibitem[{Baker(2000)}]{Baker2}
Baker, D. (2000).
\newblock A surprising simplicity to protein folding.
\newblock \emph{Nature} 405:39--42.

\bibitem[{Martinez et~al.(1998)Martinez, Pisabarro, and Serrano}]{Serrano}
Martinez, J.~C., M.~T. Pisabarro, and L.~Serrano. (1998).
\newblock Obligatory steps in protein folding and the conformational diversity
  of the transition state.
\newblock \emph{Nature Struct. Biol.} 5:721--729.

\bibitem[{Martinez and Serrano(1999)}]{Serrano2}
Martinez, J.~C., and L.~Serrano. 1999.
\newblock The folding transition state between sh3 domains is conformationally
  restricted and evolutionarily conserved.
\newblock \emph{Nature Struct. Biol.} 6:1010--1016.

\bibitem[{Riddle et~al.(1999)Riddle, Grantcharova, Santiago, Alm, Ruczinski,
  and Baker}]{Riddle}
Riddle, D.~S., V.~P. Grantcharova, J.~V. Santiago, E.~Alm, I.~Ruczinski, and
  D.~Baker. 1999.
\newblock Experiment and theory highlight role of native state topology in sh3
  folding.
\newblock \emph{Nature Struct. Biol.} 6:1016--1024.

\bibitem[{Chiti et~al.(1999)Chiti, Taddei, White, Bucciantini, Magherini,
  Stefani, and Dobson}]{Chiti}
Chiti, F., N.~Taddei, P.~M. White, M.~Bucciantini, F.~Magherini, M.~Stefani,
  and C.~M. Dobson. 1999.
\newblock Mutational analysis of acylphosphatase suggests the importance of
  topology and contact order in protein folding.
\newblock \emph{Nature Struct. Biol.} 6:1005--1009.

\bibitem[{Plaxco et~al.(1998)Plaxco, Simons, and Baker}]{Plaxco}
Plaxco, K.~W., K.~T. Simons, and D.~Baker. 1998.
\newblock Contact order, transition state placement and the refolding rates of
  single domain proteins.
\newblock \emph{J. Mol. Biol.} 277:985--994.

\bibitem[{Clementi et~al.(2000)Clementi, Nymeyer, and Onuchic}]{Clementi}
Clementi, C., H.~Nymeyer, and J.~N. Onuchic. 2000.
\newblock Topological and energetic factors: what determines the structural
  details of the transition state ensemble and "en-route" intermediates for
  protein folding ?
\newblock \emph{J. Mol. Biol.} 298:937--953.

\bibitem[{Galzitskaya and Finkelstein(1999)}]{Finkel}
Galzitskaya, O.~V., and A.~V. Finkelstein. 1999.
\newblock A theoretical search for folding/unfolding nuclei in
  three-dimensional protein structures.
\newblock \emph{Proc. Natl. Acad. Sci. USA} 96:11299--11304.

\bibitem[{Zhou and Karplus(1999)}]{Zhou}
Zhou, Y., and M.~Karplus. 1999.
\newblock Folding of a model three-helix bundle protein: a thermodynamic and
  kinetic analysis.
\newblock \emph{J. Mol. Biol.} 293:917--951.

\bibitem[{Ding et~al.(2002)Ding, Dokholyan, Buldyrev, Stanley, and
  Shakhnovich}]{Shakhnovich}
Ding, F., N.~V. Dokholyan, S.~V. Buldyrev, H.~E. Stanley, and E.~I.
  Shakhnovich. 2002.
\newblock Direct molecular dynamics observation of protein folding transition
  state ensemble.
\newblock \emph{Biophys. J} 83:3525--3532.

\bibitem[{Ozkan et~al.(2003)Ozkan, Dill, and Bahar}]{Bahar}
Ozkan, S.~B., K.~A. Dill, and I.~Bahar. 2003.
\newblock Computing the transition state populations in simple protein model.
\newblock \emph{Biopolymers} 68:35--46.

\bibitem[{Kubelka et~al.(2004)Kubelka, Hofrichter, and Eaton}]{Fast}
Kubelka, J., J.~Hofrichter, and W.~A. Eaton. (2004).
\newblock The protein folding 'speed limit'.
\newblock \emph{Curr. Opin. Struct. Biol.} 14:76--88.

\bibitem[{Gruebele(1999)}]{Fast1}
Gruebele, M. (1999).
\newblock The fast protein folding problem.
\newblock \emph{Annu. Rev. Phys. Chem.} 50:485--516.

\bibitem[{Miyazawa and Jernigan(1985)}]{MJ}
Miyazawa, S., and R.~L. Jernigan. 1985.
\newblock Estimation of effective interresidue contact energies from protein
  crystal structures: quasi-chemical approximation.
\newblock \emph{Macromolecules} 18:534--552.

\bibitem[{Clementi and Plotkin(2004)}]{Plotkin}
Clementi, C., and S.~S. Plotkin. 2004.
\newblock The effects of nonnative interactions on protein folding rates:
  theory and simulations.
\newblock \emph{Protein Sci.} 13:1750--1766.

\bibitem[{Northey et~al.(2002)Northey, Nardo, and Davidson}]{Northey}
Northey, J.~G.~B., A.~A.~D. Nardo, and A.~R. Davidson. (2002).
\newblock Hydrophobic core packing in the sh3 domain folding transition state.
\newblock \emph{Nature Struct. Biol.} 9:126--130.

\bibitem[{Di~Nardo et~al.(2004)Di~Nardo, Korzhnev, Stogios, Zarrine-Afsar, Kay,
  and Davidson}]{DiNardo}
Di~Nardo, A.~A., D.~M. Korzhnev, P.~J. Stogios, A.~Zarrine-Afsar, L.~E. Kay,
  and A.~R. Davidson. (2004).
\newblock Dramatic acceleration of protein folding by stabilization of a
  nonnative backbone conformation.
\newblock \emph{Proc. Natl. Acad. Sci. USA} 101:7954--7959.

\bibitem[{Kaya et~al.(2005)Kaya, Liu, and Chan}]{Chan}
Kaya, H., Z.~Liu, and H.~S. Chan. 2005.
\newblock Chevron behavior and isostable enthalpic barriers in protein folding:
  successes and limitation of simple go-like modeling.
\newblock \emph{Biophys. J.} 89:520--535.

\bibitem[{Honeycutt and Thirumalai(1992)}]{Thirumalai}
Honeycutt, J.~D., and D.~Thirumalai. 1992.
\newblock The nature of folded states of globular proteins.
\newblock \emph{Biopolymers} 32:695--709.

\bibitem[{Veitshans et~al.(1996)Veitshans, Klimov, and
  Thirumalai}]{Thirumalai2}
Veitshans, T., D.~Klimov, and D.~Thirumalai. 1996.
\newblock Protein folding kinetics: timescales, pathways and energy landscapes
  in terms of sequence-dependent properties.
\newblock \emph{Fold. Des.} 2:1--22.

\bibitem[{Sorenson and Head-Gordon(1999)}]{SHG3}
Sorenson, J.~M., and T.~Head-Gordon. 1999.
\newblock Redesigning the hydrophobic core of a model $\beta$-sheet protein:
  destabilizing traps through a threading approach.
\newblock \emph{Proteins: Struct. Func. Gen.} 37:582--591.

\bibitem[{Shakhnovich(1998)}]{Design}
Shakhnovich, E.~I. 1998.
\newblock Protein design: a perspective from simple tractable models.
\newblock \emph{Fold. Des.} 3:R45--R58.

\bibitem[{Evans et~al.(1983)Evans, Hoover, Failor, Moran, and Ladd}]{Isokin}
Evans, D.~J., W.~G. Hoover, B.~H. Failor, B.~Moran, and A.~J.~C. Ladd. 1983.
\newblock Non-equilibrium molecular dynamics via gauss's principle of least
  constraint.
\newblock \emph{Phys. Rev. A.} 28:1016--1021.

\bibitem[{Kabsch(1976)}]{Kabsch}
Kabsch, W. 1976.
\newblock Solution for best rotation to relate two sets of vectors.
\newblock \emph{Acta Crystallogr.} 32:922--923.

\bibitem[{Ferrenberg and Swendsen(1989)}]{Ferren}
Ferrenberg, A.~M., and R.~H. Swendsen. 1989.
\newblock Optimized monte carlo data analysis.
\newblock \emph{Phys. Rev. Lett.} 63:1195--1198.

\bibitem[{Guo and Brooks-III(1997)}]{Guo}
Guo, Z., and C.~L. Brooks-III. 1997.
\newblock Thermodynamics of protein folding: a statistical mechanical study of
  a small all-$\beta$ protein.
\newblock \emph{Biopolymers} 42:745--757.

\bibitem[{Cecconi et~al.(2001)Cecconi, Micheletti, Carloni, and
  Maritan}]{Cecconi}
Cecconi, F., C.~Micheletti, P.~Carloni, and A.~Maritan. 2001.
\newblock The structural basis of antiviral drug resistance and role of folding
  pathways in hiv-1 protease.
\newblock \emph{Proteins Struct. Funct. Genet.} 43:365--372.

\bibitem[{Micheletti et~al.(2002)Micheletti, Cecconi, Flammini, and
  Maritan}]{Cecconi2}
Micheletti, C., F.~Cecconi, A.~Flammini, and A.~Maritan. 2002.
\newblock Crucial stages of protein folding through a solvable model:
  Predicting target sites for enzyme-inhibiting drugs.
\newblock \emph{Protein Sci.} 11:1878--1887.

\bibitem[{Scala et~al.(2001)Scala, Dokholyan, Buldyrev, and
  Stanley}]{Dokholyan}
Scala, A., N.~V. Dokholyan, S.~V. Buldyrev, and H.~E. Stanley. 2001.
\newblock Thermodynamically important contacts in folding of model proteins.
\newblock \emph{Phys. Rev. E} 63:032901.

\bibitem[{Vendruscolo et~al.(2001)Vendruscolo, Paci, Dobson, and
  Karplus}]{Vendruscolo}
Vendruscolo, M., E.~Paci, C.~M. Dobson, and M.~Karplus. (2001).
\newblock Three key residues form a critical contact network in a protein
  folding transition state.
\newblock \emph{Nature} 409:641--645.

\bibitem[{Abkevich et~al.(1994)Abkevich, Gutin, and Shakhnovich}]{Abkevich}
Abkevich, V.~I., A.~M. Gutin, and E.~I. Shakhnovich. (1994).
\newblock Specific nucleus as the transition state for protein folding:
  evidence from the lattice model.
\newblock \emph{Biochemistry} 33:10026--10036.

\bibitem[{Onuchic et~al.(2000)Onuchic, Nymeyer, Garcia, Chahine, and
  Socci}]{Frustr}
Onuchic, J.~N., H.~Nymeyer, A.~E. Garcia, J.~Chahine, and N.~D. Socci. (2000).
\newblock The energy landscape theory of protein folding: Insights into folding
  mechanisms and scenarios.
\newblock \emph{Advances in Protein Chemistry} 53:87--152.

\bibitem[{Fersht(1998)}]{FershtBook}
Fersht, A.~R. (1998).
\newblock Structure and Mechanism in Protein Science: A Guide to Enzyme
  Catalysis and Protein Folding.
\newblock 3rd edition. W.~H. Freeman, USA.

\bibitem[{Settanni et~al.(2005)Settanni, Rao, and Caflisch}]{Settanni}
Settanni, G., F.~Rao, and A.~Caflisch. (2005).
\newblock $\phi$-value analysis by molecular dynamics simulations of reversible
  folding.
\newblock \emph{Proc. Natl. Acad. Sci. USA} 102:628--633.

\bibitem[{Naganathan et~al.(2005)Naganathan, Perez-Jimenez, Sanchez-Ruiz, and
  Munoz}]{baseline}
Naganathan, A.~N., R.~Perez-Jimenez, J.~M. Sanchez-Ruiz, and V.~Munoz. (2005).
\newblock Robustness of downhill folding: Guidelines for the analysis of
  equilibrium folding experiments on small proteins.
\newblock \emph{Biochemistry} 44:7435--7449.

\bibitem[{Cavalli et~al.(2005)Cavalli, Vendruscolo, and Paci}]{Cavalli}
Cavalli, A., M.~Vendruscolo, and E.~Paci. (2005).
\newblock Comparison of sequence-based and structure-based energy functions for
  the reversible folding of a peptide.
\newblock \emph{Biophys. J} 88:3158.

\bibitem[{Miller and Wales(1999)}]{Wales}
Miller, M.~A., and D.~J. Wales. (1999).
\newblock Energy landscape of a model protein.
\newblock \emph{J. Chem. Phys.} 111:6610--6616.

\bibitem[{Kaya and Chan(2000)}]{ChanK2}
Kaya, H., and H.~S. Chan. (2000).
\newblock Polymer principles of protein calorimetric two-state cooperativity.
\newblock \emph{Proteins: Struct. Func. Gen.} 40:637--661.

\bibitem[{Knott et~al.(2004)Knott, Kaya, and Chan}]{Coop}
Knott, M., H.~Kaya, and H.~S. Chan. (2004).
\newblock Energetic of protein thermodynamic cooperativity: contributions of
  local and nonlocal interactions.
\newblock \emph{Polymer} 45:623--632.

\bibitem[{Chan(2000)}]{Coop2}
Chan, H.~S. (2000).
\newblock Modeling protein density of states: Additive hydrophobic effects are
  insufficient for calorimetric two-state cooperativity.
\newblock \emph{Proteins: Struct. Func. Gen.} 40:543--571.

\bibitem[{Nymeyer et~al.(1998)Nymeyer, Garc\'ia, and Onuchic}]{Critica}
Nymeyer, H., A.~E. Garc\'ia, and J.~N. Onuchic. (1998).
\newblock Folding funnels and frustration in off-lattice minimalist protein
  landscapes.
\newblock \emph{Proc. Natl. Acad. Sci. USA} 95:5921--5928.

\bibitem[{Kwiecinska and Cieplak(2005)}]{Cieplack}
Kwiecinska, J.~I., and M.~Cieplak. (2005).
\newblock Chirality and proteins folding.
\newblock \emph{J. Phys.: Condens. Matter} 17:S1565--S1580.

\end{thebibliography}
\end{document}